%% file: re-flat-revised.tex
\documentclass[final,1p]{elsarticle}

\journal{Journal of \LaTeX\ Templates}

\journal{Annals of Physics}

\usepackage{graphicx}
\usepackage{dcolumn}
\usepackage{amsmath,amssymb}
\usepackage[]{amsfonts}
\usepackage[]{latexsym}
\usepackage[]{float}
\usepackage[all]{xy}
\usepackage{bm}
\usepackage{ulem}
\usepackage{xcolor}

\usepackage{bigdelim,multirow}

\newcommand{\cin}[1]{{\color{red}#1}}
\renewcommand{\cin}[1]{{\color{black}#1}}

\input{mydef}

\begin{document}

\begin{frontmatter}

  \title{Revisiting Flat bands and localization}

\author{Yasuhiro Hatsugai}
\address{Department of Physics, University of Tsukuba, 1-1-1 Tennodai, Tsukuba, Ibaraki 305-8571, JAPAN}

\begin{abstract}
  Flat bands imply lack of itinerancy due to some constraints
  that, in principle,  results
  in  anomalous behaviors with randomness.
  By a molecular orbital (MO) representation of  the flat band systems,
  random MO models are introduced where 
  the degeneracy due to the flat bands is preserved even with randomness.
The zero modes of the chiral symmetric system with sublattice imbalance belong to the class.
  After explaining the generic flat band construction by MOs,
 several examples are discussed with numerical demonstration
  as  \cin{sawtooth} lattice in one dimension and 
  hyper-Pyrochlore lattice in any $d$-dimensions that extends
  the Kagome ($d=2$) and Pyrochlore ($d=3$) lattices to general dimensions. 
\end{abstract}

\begin{keyword}
  \texttt{Flat bands\sep molecular orbitals \sep Kagome \sep Pyrochlore \sep randomness}
\end{keyword}

\end{frontmatter}


\section{Introduction}
Massless Dirac fermions with singular dispersion
are sources of non-trivial topology and get focused substantially
in relation to topological phases and  Anderson localization as well.
Flat bands are another singular dispersion and have a long history of studies. 
Any projection operator of an eigen state in momentum space has degenerate eigen values (0 and 1),
that gives a trivial  example of the flat band.
However, constructing a tight-binding Hamiltonian with strictly short range hopping is not
trivial. 
A tight-binding Hamiltonian of a simplified $sp^3$ network of an amorphous solid
is one of the oldest examples of the flat band systems\cite{Weaire71}.
The origin of the flat bands in the model was  clarified based on
the molecular orbital (MO) construction\cite{Hatsugai11ZQ} 
and the MO representation is
applied to the simplified model of silicene as a two-dimensional analogue\cite{Hatsugai15WT}.
The systems with flat bands also appeared occasionally in  various studies from quite
different points of view such as
fermion doubling\cite{Dagotto86}, ferromagnetism\cite{Mielke91,Tasaki92,Katsura10} and
\cin{ferrimagnetism\cite{Lieb89}}. \cin{The flat band is also observed experimentally in
   frustrated Kagome metal\cite{Kang20}}.
The flat bands of local Hamiltonian are due to some local constraints,
which imply  possible non trivial behaviors in Anderson localization
such as multifractality\cite{Goda06, Chalker10}.
One of the well known origin of the flat bands is imbalance of the sublattices in a chiral symmetric
system.
Appearance of localized boundary modes of graphene near the Zigzag edges belong also to the class.
A hopping Hamiltonian on the Penrose tiling \cin{is} also in this class
since the Penrose tilling is bipartite and there exist sublattices imbalance\cite{Kohmoto86p}.
Although the system is aperiodic, the Hamiltonian hosts macroscopic degeneracy (ring states) due to
the constraint. This is an example of "flat band'' without translational symmetry.

Recently we are proposing a general scheme to construct a strictly local Hamiltonian with flat bands
\cite{Hatsugai11ZQ,Hatsugai15WT,Mizoguchi19flat,Mizoguchi20sys}.
It includes the above chiral symmetric sublattice imbalance case.
The construction is independent of the translational symmetry and can be applied to
random systems.
We are here introducing such a class of systems
as a random MO model and 
present some of numerical results as a demonstration.
As for the numerical evaluation of the probability
density $p_i$ and its correlation function $C_{ij}$,
we do not need diagonalizaton.
Also the numerical studies can be done using a sparse matrix technique since the MO is local. It enables us to
  treat large systems.

  \section{Molecular Orbital representation}
Let us first define $M$ molecular orbitals (MOs) $|i )$,
($i=1,\cdots,M$) as
\begin{align*}
  | i ) = \sum_{n=1}^N| n \rangle \psi  _{n,i}=|{\cal A} \rangle \psi_i,\quad 
|{\cal A} \rangle  =(|1 \rangle ,\cdots,|N \rangle ),
\end{align*}
where
$\psi_i$ is an $N$-dimensional column vector of the $i$-th molecular orbital
and $| {\cal A} \rangle $ is a set of 
orthonormalized atomic basis,
$\langle n|m \rangle = \delta _{nm}$,
$\langle {\cal A}| {\cal A} \rangle =I_N $ ($N$ is the number of atoms).

A Hamiltonian written by the $M$ molecular orbitals (MOs) is 
\cite{Hatsugai11ZQ,Hatsugai15WT,Mizoguchi19flat,Mizoguchi20sys}
\begin{align*}
{\cal H}     &= \sum_{i,j}^M| i )h_{i,j}(j|=|{\cal M} ) {h} ({\cal M} |
  =|{\cal A} \rangle {H} \langle{\cal A} |
  \\
H    &= \Psi h \Psi ^\dagger ,
\end{align*}
where
$H\in M(N,N)$, 
$ {h} \in M(M,M)$
\footnote{ $M(n,m)$ is a set of $n\times m$ matrices.}
and $ h_{i,j}=\{ h \}_{i,j}$,
is a hopping between the molecular orbitals $|i)$ and $|j)$.
Here $ |{\cal M} ) $ is a set of the MOs' basis 
\begin{align*}
  |{\cal M} ) &= (|1),\cdots,|M))= |{\cal A}  \rangle \Psi,\quad
  \Psi=(\psi_1,\cdots,\psi_M)\in M(N,M).
\end{align*}
We assume that
 $\det h\ne 0$ and
the MOs $\psi_i$, $i=1,\cdots M$
are linearly independent 
(${\rm rank}\,\Psi=M$)
\footnote{If $\det h=0$, one can reduce the independent MOs.
}.
It implies the overlap matrix
${\cal O}=({\cal M}|{\cal M})= \Psi^\dagger\Psi$
is positive definite ($\det O\ne 0$).

If $N>M$, this Hamiltonian ${\cal H} $ has \cin{at least} $N-M$ degenerate zero modes 
since 
the zero energy condition, 
$( {\cal M} |\varphi \rangle =0$ for $|\varphi \rangle =|{\cal A} \rangle \varphi $,
gives $M$ linear \cin{conditions} for $N$-dimensional vector $\varphi$.
\footnote{
It also obeys from a simple algebra for the secular equation
$\det_N( \lambda I_N-H)=\det_N( \lambda I_N-\Psi h\Psi ^\dagger )=
\lambda ^{N-M}\det_M( \lambda I_M- h {\cal O}  )$
\cite{Hatsugai11ZQ,Mizoguchi19flat,Mizoguchi20sys}.
}
\footnote{
When the system is periodic,
there exists flat bands. One may also consider $H$ as a Hamiltonian in momentum representation and the label $n$ as an index of
the atomic species within the unit cell in momentum space. Then the zero energy flat band is $N-M$ fold degenerated.
}
\footnote{ This MO construction of the zero mode flat bands reminds us of
the Maxwell relations for stability/instability of mechanical systems\cite{Maxwell,Vitteli15}.
}
This is true even when $\Psi$ is random, which we propose as a random molecular orbital model\cite{Hatsugai11ZQ,HK17,Kohda17}.

By using an orthnormalized basis of the MOs,
\begin{align*}
  | \bar {\cal M} ) &=  |  {\cal M} ) {\cal O} ^{-1/2},
  \  ( \bar {\cal M} |\bar M ) = I_M,
\end{align*}
the projection to the non-zero energy states, ${\cal P}_1  $ is written as
\begin{align*}
    {\cal P} _1 &= |\bar {\cal M})  ( \bar{\cal M} |
    = |{\cal A}\rangle\Psi {\cal O} ^{-1} \Psi ^\dagger \langle {\cal A} |.
\end{align*}
Now we have an orthogonal decomposition of the one-particle Hilbert space as
\begin{align*}
  1 &= {\cal P} _{0}+{\cal P} _{1},  \ \
    {\cal P} _i{\cal P} _j =  \delta _{ij} {\cal P} _i,
\end{align*}
where ${\cal P}_1 $ and  ${\cal P}_0 $ are
projections to the non-zero/zero energy states respectively
($    {\cal H}{\cal P}_1 =  {\cal H}$
and
$      {\cal H}{\cal P}_0 = 0$).
Then the complementary projection to the zero modes is written as
\begin{align*} 
  {\cal P} _0 &= 1-{\cal P}_1= |Z \rangle \langle Z | ,\ \langle Z|Z \rangle =I_{N-M} 
  \\
  | Z \rangle &= (|\varphi_1 \rangle ,\cdots,|\varphi_{N-M} \rangle),\ \langle M|\varphi_i \rangle =0
  \\
  |\varphi_\ell \rangle &= | {\cal A} \rangle \varphi_\ell,\ H\varphi_\ell=0,
\end{align*}
where
$|\varphi_\ell \rangle $, $\ell=1,\cdots,N-M$ are \cin{orthonormalized} states of the zero mode,
\cin{ $\langle  \varphi_\ell|\varphi_{\ell ^\prime }\rangle =\delta _{\ell\ell ^\prime }$}.
  
\cin{To discuss properties of localization of the wave function $\varphi_\ell$ of the zero
  energy, the probability distribution $p_i^\ell=C^{\ell}_{ii}=|\varphi_{i,\ell}|^2$ and
  the correlation function $C^{\ell}_{ij}=\varphi_{i,\ell}\varphi_{j,\ell}^*$ can be useful.
  However, 
  the exact degeneracy of the flat band implies that each $\varphi_\ell$ is not uniquely defined and
  one may take another set of the zero energy states,
  $|Z ^\prime \rangle = (|\varphi_1 ^\prime \rangle ,\cdots,|\varphi_{N-M}^\prime \rangle$.
  Of course $|Z \rangle $ and $| Z ^\prime \rangle $ are unitary equivalent
  as $| Z ^\prime  \rangle =|Z \rangle U$, $U\in U(N-M)$.
  Then what is invariant (independent of the transformation $U$)
  is an average or a trace over the degenerate states
defined  as
  }
  \begin{align*}
    C_{ij} &= \cin{\sum_{\ell=1}^{N-M} C^\ell_{ij}=}
    \sum_{\ell=1}^{N-M}    \varphi_{\cin{i,\ell}}    \varphi_{\cin{j,\ell}}^*
=    
    \langle i| {\cal P} _0 | j \rangle 
    =
           \delta _{ij}-\tilde C_{ij}
   \\
   \tilde C_{ij} &= (\Psi {\cal O} ^{-1} \Psi ^\dagger )_{ij}.
  \end{align*}
  \cin{
    Its invariance is clear
    by ${\cal P}_0=|Z \rangle   \langle   Z|=|Z \rangle   \langle   Z ^\prime|  $.
    }
  The local probability of the site $i$ for the zero mode space
  defined as $p_i=C_{ii}/M$, ($ \sum_i p_i=1$) is \cin{the average of the
    local probability distribution of $p_i^\ell$. This is the only well-defined quantity
    that is independent of the choice of the states.
    One may wonder effects of the randomness might be washed out by the average.
    However it is not the case as shown in the concrete examples discussed
    in the next section.
  }
\cin{Since the projection into the zero energy space ${\cal P}_0 $ } is independent of the hopping between the MOs, $h_{ij}$, 
we assume  $h=I_M $ without loss of generality as far as the degenerate zero energy space is concerned.

  Note also that $C_{ij}$ is also written as a correlation function of the filled zero modes
$    |z\rangle _F $ of the second quantized fermions as 
  \begin{align*}
    C_{ij} &= {_F\langle} z| c_j ^\dagger  c_i | z \rangle _F 
   \\
   |z\rangle _F &= \prod_{\ell=1}^{N-M} \bm{c} ^\dagger \varphi_\ell | 0 \rangle\cin{_F},
  \end{align*}
  \cin{
    since ${_F \langle }z| c_i c_j ^\dagger | z {\rangle _F}=\delta _{ij}- \langle i| {\cal P} _0| j \rangle$ 
  }
  where   $c_i$, ($i=1,\cdots N$) is a fermion annihilation operator ($\acmt{c_i}{c_j ^\dagger }=\delta _{ij}  $),
$\bm{c} ^\dagger =(c_1,\cdots,c_N) $
  and $|0 \rangle\cin{_F} $ is their vacuum ($c_i |0 \rangle\cin{_F} =0$). \cin{
    See appendix A.

  }

  \section{Examples}
  \subsection{Sawtooth lattice in 1D}
  
\begin{figure}[h]
\includegraphics[width=150mm]{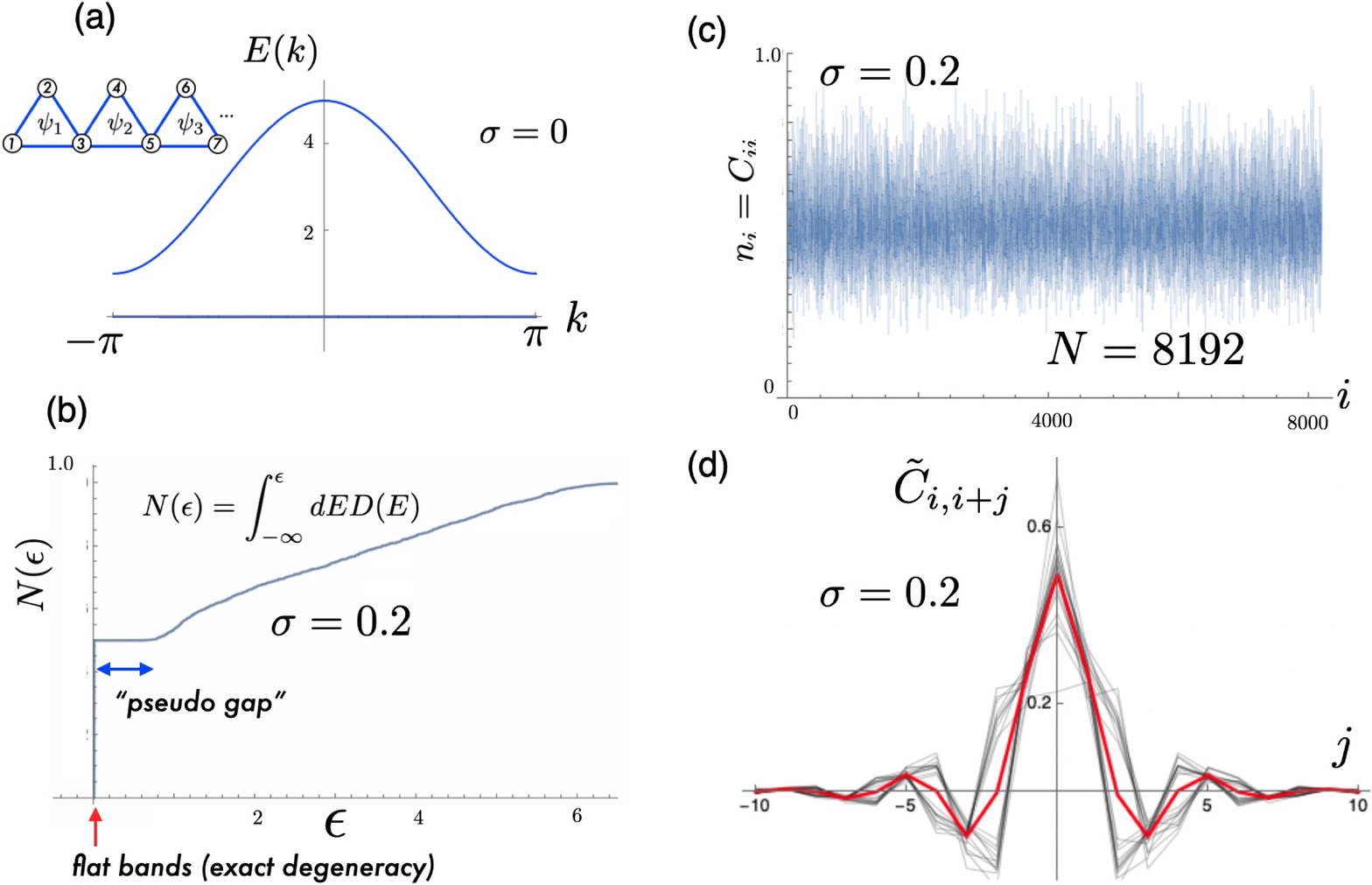}
\caption{\label{fig:sw} 
  (a): Sawtooth lattice and its energy dispersion for $a=b=c=1$.
  (b) Integrated density of states $N(\epsilon )$ ($\sigma =0.2$)
  (c) "Charge'' $n_i=C_{ii}$ of the zero energy space ($N=2M=8192$) with periodic boundary condition.
  Normalized probability density $p_i=n_i/M$. 
  (d) "Correlation function'' $\tilde C_{i,i+j}$ of the zero energy space.
  The black lines are for 21 different $i$'s and
  the red one is their average.
}
\end{figure}

The simplest example with a flat band is a one-dimensional \cin{sawtooth} lattice ( Fig.\ref{fig:sw}(a)) This model has been discussed in many papers (See.\cite{Kuno20st}, for example).
  The Hamiltonian is written by $M$-MO,
  $|j )=|\xi_j  \rangle \xi_j^1+|\eta_j \rangle \eta_j+|\xi_{j+1} \rangle \xi_j^2 $, ($j=1,\cdots,M)$
  for an $N=2M$ site system as
  $H=\sum_{j=1}^M | j )(j|$ where $|\xi_j \rangle $ and $|\eta_j \rangle $ are orthonormalized
  atomic states and 
  $\xi_j^1,\xi_j^2,\eta_j  \in\mathbb{C}$ are arbitrary.
  We assume periodic boundary condition $|\xi_{M+1}\rangle =|\xi_1 \rangle $ for simplicity.
  Since the Hamiltonian is written by $M$ MOs and the number of the atomic sites is $N=2M$, it has $N-M=M$
  degenerate zero energy states.
  If the system is translational invariant as $\xi_j^1=a,\xi_j^2=b,\eta_j=c$,
  one has $H =\sum_{k_n} |k_n ) ( k_n|$,
  $ |k_n )=(|\xi_{k_n} \rangle,|\eta_{k_n} \rangle )\psi_{k_n}$
$^t\psi_{k_n}=  (a + c e^{ik_n}, b)  $ in
  the momentum representation
  $|\xi_j ) = M ^{-1/2}\sum_{k_n} | \xi_{k_n} ) e^{i k_n j} $,
  $|\eta_j ) = M ^{-1/2}\sum_{k_n} | \eta_{k_n} ) e^{i k_n j} $,
  $k_n=\frac {2\pi n}M, n=1,\cdots,M$.
  Now its non-zero energy band is given by $\psi_{k_n} ^\dagger \psi_{k_n}=|a + b e^{ik_n}|^2 + |c|^2$ and
  the rest is the flat band at the zero energy.

  We have calculated probability distribution $p_i$ and its
  correlation function $C_{ij}$ for the random MO sawtooth model in Fig.\ref{fig:sw}(b)-(d).
  The random MOs are defined by $\xi_i^{1,2}=1+X_i^{1,2} $ and $\eta_i=1+X_j $ where $X_i^{1,2}$
  and $X_i$ are random variables of the normal distribution with variance $\sigma^2 $ and its mean $0$.
  In the Fig.\ref{fig:sw}(b), the integrated density of states is shown for $\sigma =0.2$.
  The charge $n_i$ is shown for $\sigma =0.2$ for $N=8192$ system in Fig.\ref{fig:sw}. It is very spiky
  and behaves randomly in a microscopic scale.
  We have also calculated  inverse participation ratios (IPR)\cin{,} $I=\sum_i p_i^4$, ($p_i=n_i/M$, $\sum_i p_i=1$)
  for different system sizes ($N=8$ to  $N=8192$). It behaves ${\cal O}( 1/N)$ that suggests
  the space is  extended \cin{since it is an average of the {\rm IPR} of
  the degenerated zero energy states}.
  Also the correlation function $\tilde {C}_{i,i+j}$ is plotted in Fig.\ref{fig:sw}(d) averaged over several $i$'s.
  It suggests the average $\tilde {C}_{ij} $ is exponentially localized, that is,
  the zero energy space is short-ranged and its localization length is roughly several lattice spacing.
  It implies that the zero mode space is intrinsically localized due to the local nature of the MO
  within microscopic space and the macroscopic degeneracy makes the whole zero mode space extended as is shown
  in the IPR.
  
\subsection{Hyper-Pyrochlore lattice in $d$-dimensions}
\begin{figure}[h]
\begin{center}
  \includegraphics[width=110mm]{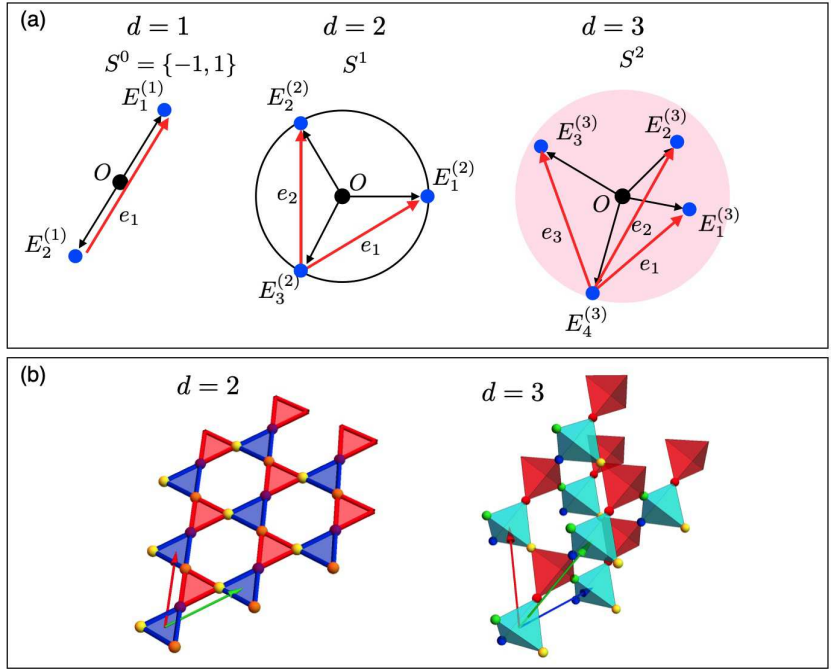}
  \end{center}
  \caption{\label{fig:unit}
    (a)Unit vectors $e_i$, $j=1,\cdots,d$ for
    a series of hyper-Pyrochlore lattices\cite{Hatsugai11ZQ}.
    Equivalent $(d+1)$ points $E^{(d)}_{j}$, $j=1,\cdots,d+1$ are distributed on the Sphere $S^{d-1}$.
(b) Several lattice points and its unit vectors are shown for $d=2$ (Kagome) and $d=3$ (Pyrochlore).
}
\end{figure}
We have also proposed a series of systems with flat bands in $d$-dimensions
as a hyper-Pyrochlore lattice\cite{Hatsugai11ZQ}.
This is dual to the $d$-dimensional graphene that is an extension of 4-$d$ graphene\cite{Creutz08}.
Energy bands of the hyper-Pyrochlore are given by
the (shifted) bands of $d$-dimensional graphene with extra
$(d-1)$-fold degenerated flat bands  at zero energy\cite{Hatsugai11ZQ}.
As the $d$-dimensional graphene is a minimum model for the massive/massless Dirac fermions in $d$-dimension,
the hyper-Pyrochlore lattice is also minimum as a system with flat bands in general dimension.

Let us start from $(d-1)$-dimensional sphere $S^{d-1}$ (See Fig.\ref{fig:unit}(a))
and $d+1$ points, $E^{(d)}_1,\cdots,E^{(d)}_{d+1}$ which are equivalently distributed on the unit sphere
($E^{d}_j\cdot E^{(d)}_{j ^\prime }=-d ^{-1} $, $j=1,\cdots,d+1$).
\footnote{They are recursively constructed from $d=1$ by
\begin{align*}
  E^{(d)}_j &=
  \mchss
      {\alpha _d E_j^{(d-1)}+\beta _d E^{(d)}_{d+1}}
      {j=1\cdots,d}
  {^t(0,\cdots,0,1_d,0,\cdots)}{j=d+1},
\end{align*}
where $\alpha _d=\sqrt{1-d^{-2}}$ and $\beta _d= - d ^{-1} $.
Here we have assumed the system is embedded in sufficiently large dimension.
}
Then the unit translations,
$e_1,\cdots,e_d$, are given by $e_j=e^{(d)}_j=E^{(d)}_j-E^{(d)}_{d+1}$.
The unit cell of the hyper-Pyrochlore lattice includes $d+1$ atoms which
locate at 
 $-e_1/2,\cdots,-e_{d+1}/2$, ($e_{d+1}=0$ is supplemented).

\begin{figure}[h]
\includegraphics[width=150mm]{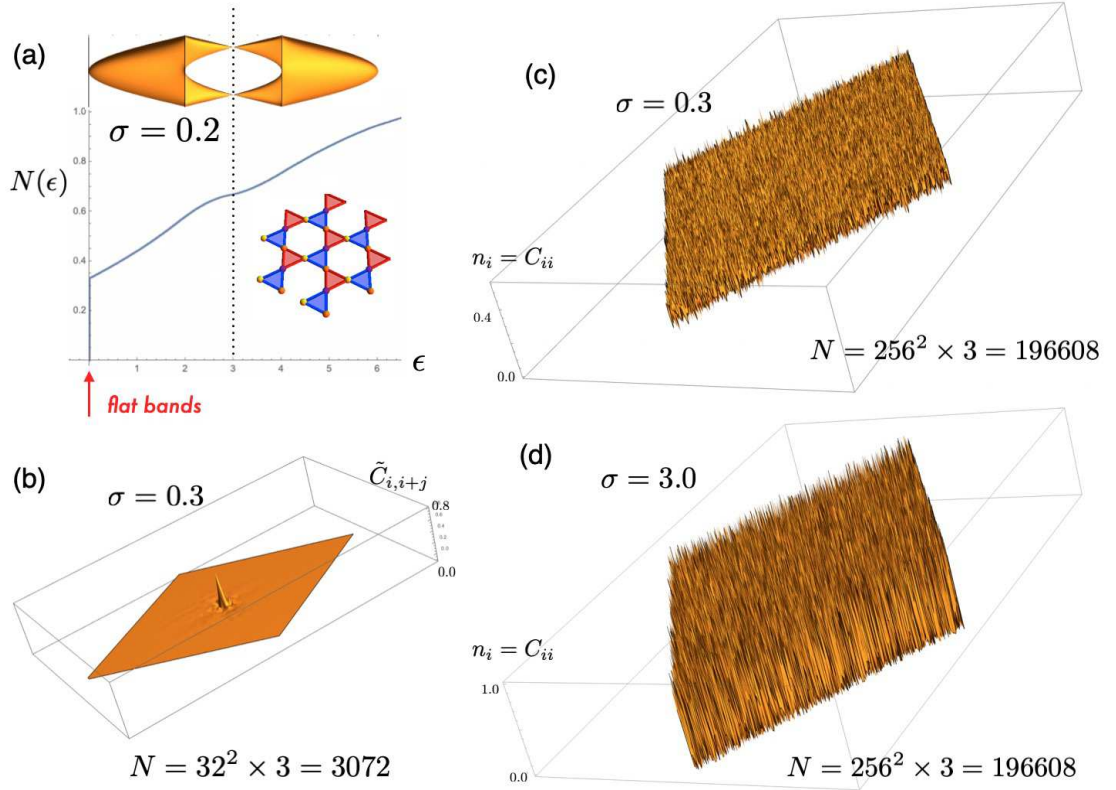}
\caption{\label{fig:kagome} 
  (a) Integrated density of states for Kagome lattice without randomness ($\xi^j_r=\eta^j_r=1$).
  (b) "Correlation function'' $\tilde C_{i,i+j}$ of the zero energy space ($N=3\times 32^2$). 
  (c) "Charge'' $n_i=C_{ii}$ of the zero energy space ($N=3M=3\times 256^2$)
  with periodic boundary condition. Normalized probability density $p_i=n_i/M$. ($\sigma =0.3$)
  (d) "Charge'' $n_i=C_{ii}$ of the zero energy space ($N=3M=3\times 256^2$)
  with periodic boundary condition. Normalized probability density $p_i=n_i/M$. ($\sigma =3.0$)
}
\end{figure}
  
  With periodic boundary condition of the linear size $L$,
  $(d+1)L^d$ atomic positions of the system with $L^d$ unit cells
  are
  \begin{align*}
r_j &=    r-e_j/2,\ (j=1,\cdots,d+1),
  \end{align*}
  where
  $    r = n_1 e_1+\cdots +n_d e_d$ and $ n_j=1,\cdots,L$, ($j=1,\cdots,d$). See Fig.\ref{fig:unit}(b).
    The Hamiltonian of the hyper-Pyrochlore is written as
  \begin{align*}
H &=     \sum_{r}( |B_r )(B_r|+ |R_r )(R_r|).
  \end{align*}
Two MOs, $|B_r )$ and $|R_r)$ at the unit cell $r$ are defined as
  \begin{align*}
    |B_r ) &= \sum_{j=1}^{d+1}| r_j  \rangle \xi_r^j
    \\
    |R_r ) &= \sum_{j=1}^{d+1}| r  +e_j \rangle \eta_r^j    \cin{,}
  \end{align*}
  where $| r _j \rangle $ and $|r_j+e_j \rangle $ are orthonormalized atomic states at the sites.
  The weights (coefficients) within the MO, $\xi_r^j, \eta_r^j\in \mathbb{C}$, can be any.
  As for the numerical demonstration,  we take
  $\xi_r^j=1+X_r^j$ and $\eta_r^j=1+Y_r^j$ where $X_r^j$ and $Y_r^j$ are random variables of the normal distribution
  with variance $\sigma ^2$ and its means $0$.
  Since the number of the atomic sites is $N=(d+1)L^d$ and that of the MOs is $M=2L^d$, the system
  has $N-M=(d-1)L^d$ zero energies. If the system is translationally invariant, that is,
  $\xi_r^j$ and   $\eta_r^j$ are $r$-independent, the system has (at least) $(d-1)$-fold degenerate
  flat bands (at the zero energy).
  \cin{
    In a translationally invariant clean system, the Hamiltonian in momentum space is
    written by the MOs in momentum space.
    Then the flat band  is give by the momentum independent
    zero mode(see footnote in page 2).
    In this case, the MOs are momentum dependent and may not be always linearly
    independent. If the MOs are not independent, 
    $\det {\cal O} =0$, it causes extra degeneracies such as
    band touching\cite{Bergman08,Hatsugai11ZQ,Mizoguchi19flat}.
  }

  Numerical results for the case $d=2$, that is, the random MO Kagome lattice is shown in Fig.\ref{fig:kagome}.
  They are comparable to the results of the one-dimensional \cin{sawtooth} lattice.
  In Fig.\ref{fig:kagome}(a), the integrated density of states
  for the random MO Kagome model is shown ($\sigma =0.2$).
  The Dirac cones for the clean case is also shown for  reference.
  The ``correlation  function'' $\tilde C_{i,i+j}$ of the zero energy space is shown in Fig.\ref{fig:kagome}(b) for a rather
  small system  ($N=3\times 32^2$) $i$ is at the center of the system. It also shows the correlation function
  is short-ranged over  several lattice spacing.
  Charge distributions/probability distribution $n_i$ are shown in Fig.\ref{fig:kagome}(c),(d) for $\sigma =0.3$ and
  $\sigma =3.0$ respectively ($N=3\times 256^2$).
  They show the distribution is very spiky over several to 10 lattice spacing.
  We have also calculated IPR, $I=\sum_i p_i^4$, ($p_i=n_i/M$, $\sum_i p_i=1$) for $N=3\times L^2$, $L=2^4$ to $2^7$. \cin{This is the average of the {\rm IPR}'s over the degenerate zero energy states
    this is unitary invariant 
  }. 
  They behave ${\cal O}(1/L^2) $ that is consistent with the extended states.
  It suggests the zero mode space is microscopically quite spiky but extended globally.
  It is similar to the results of the \cin{one-dimensional} sawtooth lattice.
\footnote{
  When one uses the method in Ref.\cite{Jansen89}  for calculating a singularity spectrum (multifractal analysis),
  after averaging over a box of the size $\ell$, taking an extrapolation $\ell\to 0$ is needed.
  Using the average over  the box size $\ell$ up to
   several lattice spacing,
   numerical multifractal analysis of the wave
   functions of the random MO Kagome lattice has been performed
   by exact diagonalizations\cite{Kohda17}.
   It suggests multifractality of the flat band wave functions
   in a microscopic scale (over several lattice spacing).
   To obtain  conclusive results, further  analysis is needed.
  }
  \cin{Similarly} to the results of the \cin{sawtooth} lattice, the zero mode space of the random MO Kagome lattice is
  intrinsically  short-ranged (this is also true without any randomness).
  It is consistent with the spiky behavior of the probability distribution within a microscopic scale.
  However, due to the macroscopic degeneracy,
  the space as a whole is extended as shown by the IPR.

  \cin{
    At the mobility edge of the Anderson localization as the critical point,
    one may naturally expect multifractal behaviors of the wave function
    since the localization length diverges and the system is scale free.
    Away from the critical point, such  a  multifractal behavior is cut off
    at the localization length. It implies that 
    that the wave function of localized states
    is multifractal-like up to the localization length.
    Even for the present Kagome flat band system,
    one may expect standard Anderson localization except the flat bands.
    As discussed, the zero mode flat band space, that is, the average of
    the flat bands are complementary to the sum  of them those are localized and
    their localization centers are randomly chosen. The spiky structure observed in $C_{ij}$ is reflecting this multifractal-like behavior within the localization length.    
  }
  
\subsection{Bipartite lattice}
If the lattice sites are divided into two sublattices A and B and the hopping matrix elements are only non-zero
between them, the lattice is bipartite and the corresponding Hamiltonian written in the atomic basis is chiral symmetric.
By the choice of the MO in Ref.\cite{Mizoguchi19flat}, 
it is written by the MOs as 
\begin{align*}
  {\cal H} _{C} &= \sum_{a\in A,b\in B}
  ( | a \rangle_A t_{ab}{_B\langle} b|+| b \rangle_B t_{ab}^*{ _A\langle} a|)
  \\
  &= | {{\cal A} } \rangle  D_{AB} \langle {\cal B}  |+| {\cal B}  \rangle  D_{BA} \langle {\cal A}  |
  \\
  &=| {\cal A},{\cal B}   \rangle H_C \langle {\cal A} ,{\cal B} |\\
  H_C &= \mmat{O_{AA}}{D_{AB}}{D_{BA}}{O_{BB}}
  \\
  |{\cal A},{\cal B} \rangle &= (|{\cal A} \rangle ,|{\cal B} \rangle), \ \
  | {\cal A}  \rangle  =(|1 \rangle_A ,\cdots,|N_A \rangle _A), \ \
    |{\cal B}   \rangle  =(|1 \rangle_B ,\cdots,|N_B \rangle _B),
\end{align*}
where $D_{BA}=D_{AB} ^\dagger\in M(N_A,N_B) $, $\{D_{AB}\}_{ab}=t_{ab}$
and  $H_C$ is chiral symmetric $\acmt{H_C}{\Gamma }=0$ with $\Gamma={\rm diag}\, (I_A,-I_B)  $.
It is also written by $M=2N_B$ MOs, $\Psi$ as \cite{Mizoguchi19flat}
\footnote{ $O_{\alpha \beta }\in M(N_\alpha, N_\beta )$ is a zero matrix ($\alpha , \beta \in \{A,B\}$).}
\begin{align*}
  {\cal H} _C &= |{\cal M} ) h (\cin{{\cal M}} |\cin{,} \
  |  {\cal M} ) = (|{\cal A} \rangle D_{AB} ,\ |{\cal B} \rangle) = |{\cal A},{\cal B} \rangle \Psi
  \\
  \Psi &= \mmat
       {D_{AB}}{O_{AB}}
       {O_{BB}}{I_{N_B}} \in M(N_A+N_B,2 N_B)\cin{,}\\
       h &=  \mmat{O_{BB}}{I_{N_B}}{I_{N_B}}{O_{BB}}\in M(2N_B,2N_B).
\end{align*}
Since the number of total atomic sites is $N=N_A+N_B$, the Hamiltonian has $Z=N-M=N_A-N_B$ zero modes
if $N_A> N_B$. (The case, $N_B>N_A$, is discussed similarly.)
It implies that the chiral symmetric (random) system is
regarded as the (random) MO model.
One of the historical examples of the flat band due to this mechanism
is the ring states of the Penrose tiling\cite{Kohmoto86p}
\cin{and discussed from a generic point of view\cite{Surtheland86}.}
Another typical example is the zero energy state of the random Dirac fermions\cite{Hatsugai97},
where $N_A-N_B=1$ and the zero mode wave function is
discussed in relation to its multifractality.

\section{Discussion}
As has been discussed  previously\cite{Goda06, Chalker10},
the states near the flat band energy become singular by randomness
and their multifractality has been implied.
In the random MO models presented in examples of the sections 2 and 3, macroscopic degeneracy of
the flat bands is preserved even with randomness.
This \cin{macroscopic} degeneracy  makes the situation unclear although
the zero mode space
obtained numerically is very local and singular in a \cin{microscopic scale}.
It is also consistent with the numerical calculation in Ref.\cite{Hatsugai97} where the zero mode is not degenerated ($Z=N-M=1$).
Then introducing  extra MOs
associated with the randomness can be interesting. If one introduces
macroscopic $M ^\prime $ extra random MOs ($M+M ^\prime \lesssim N$), the dimension of the zero energy space can be
${\cal O}(1) $.
Also the MO representation are useful for diagonalization of random systems within the projected flat band subspace.

  \section{Acknowledgment}
  We thank discussion
  with T. Mizoguchi and T. Kuroda.
  This work is supported in part by the JSPS KAKENHI, Grant No. JP17H06138.

  \cin{

    \appendix
  
\section{Fermion correlation function}

Let us here discuss fermion correlation functions for a
filled fermi state.
We label the Fock state $|\Psi \rangle _F$ specified by the multiplet,
$\Psi\in M(N,M)$,
of the dimension $M$ for the dimension $N$ Hilbert space as
\begin{align*} 
|\Psi \rangle_F &= 
\prod_{\ell=1}^M (  {c} ^\dagger \psi_\ell )| 0 \rangle_F
=
(  {c} ^\dagger \psi_1 )\cdots(  {c} ^\dagger \psi_M )
| 0 \rangle_F=
\prod_{\ell=1}^M (c_j ^\dagger \psi^j_\ell )| 0 \rangle_F
\\
\Psi &=  (\psi_1,\cdots,\psi_M),\
\psi_\ell =
\mvecthree{\psi_{1,\ell}}{\vdots}{\psi_{N,\ell}},\ 
{c} =\mvecthree{c_1}{\vdots}{c_N},
\end{align*} 
where $|0 \rangle _F$ a vacuum, $c_i | 0 \rangle _F=0$, and
the ordering of the fermions are implicitly assumed as shown above.
The overlap of such two fermionic states $|\Phi \rangle_F$ and
$|\Psi \rangle _F$  
is directly evaluated as
\begin{align*} 
{_F\langle} \Psi |\Phi {\rangle_F} &=  \det\nolimits_M \Psi ^\dagger \Phi.
\end{align*} 

Using this, the  correlation function, 
$g^U_{ab} = {_F\langle} \Psi |c_a c_b ^\dagger |\Phi {\rangle_F} $,
is written as
\begin{align*} 
g^U_{ab} &={_F\langle} \Psi |c_a c_b ^\dagger |\Phi \rangle_F = 
\det\nolimits_{M+1} \Psi_a ^\dagger \Phi_b
\end{align*} 
where
$ \Phi_a = (e_a,\Phi)\in M(N,M+1)$,
$ \Psi_b = (e_b,\Psi)\in M(N,M+1)$
are obtained from $\Phi$ and $\Psi$ 
by adding column
vectors $e_a$ and  $e_b$  respectively
where $\{e_{i}\}_j = \delta_{ij}$.

Using a formula for the determinant
\footnote{
\cin{
As for
$A\in M(N,N)$,
$B\in M(N,M)$,
$C\in M(M,N)$,
$D\in M(M,M)$ and $\det_M D\ne 0 $, the following relation holds,
\begin{align*}
  \det\nolimits_{N+M}\mmat{A}{B}{C}{D}
  &=
\det\nolimits_{N+M}
\mmat{I_N}{-B D ^{-1}  }{O_{MN}}{I_M }
\mmat{A}{B}{C}{D}
=
\det\nolimits_{N+M} \mmat{A-B D ^{-1} C}{O_{NM}}{C}{D}
\\
&=
\det\nolimits_N (A-B D ^{-1} C)\det\nolimits_M{D}
\end{align*}
}
},
it is calculated as
\begin{eqnarray*}
g^U_{ab} &=& \det\nolimits_{M +1}
\mvec
{e_a ^\dagger }{\Psi ^\dagger }
(e_b,\Phi)
=
\det\nolimits_{M+1} \mmat
{e_a ^\dagger e_b}{e_a ^\dagger \Phi  }
{\Psi ^\dagger e_b}{\Psi^\dagger \Phi  }
\\
&=& 
\det\nolimits_{M+1} \mmat
{e_a ^\dagger e_b
-
e_a ^\dagger \Phi (\Psi^\dagger \Phi) ^{-1} \Psi ^\dagger e_b
}{0  }
{\Psi ^\dagger e_b}{\Psi^\dagger \Phi  }
\\
&=&
\det\nolimits_{1}[ e_a ^\dagger e_b
-
e_a ^\dagger \Phi (\Psi^\dagger \Phi) ^{-1} \Psi ^\dagger e_b
]
\det\nolimits_{M} (\Psi ^\dagger \Phi)
\\
&=&
\det\nolimits_{1}[ e_a ^\dagger \big[
I_N
-
\Phi (\Psi^\dagger \Phi) ^{-1} \Psi ^\dagger \big]e_b
]
\det\nolimits_{M} (\Psi ^\dagger \Phi)
\\
&=& 
\big[I_N-\Phi (\Psi^\dagger \Phi) ^{-1} \Psi ^\dagger \big]_{ab} \det\nolimits_{M} (\Psi ^\dagger \Phi).
\end{eqnarray*} 
It implies
\begin{align*} 
  g_{ab} &\equiv 
  \frac{
    {_F\langle}\Psi|c_a c_b ^\dagger|\Phi{ \rangle _F}
    }
       {
         {_F\langle}\Psi|\Phi{ \rangle _F}
       }
       = 
      g^U_{ab }/\det\nolimits_M(\Psi^\dagger  \Phi)  = [g]_{ab}
\\
g &= I_N - \Phi(\Psi ^\dagger \Phi ) ^{-1} \Psi ^\dagger 
\\
\tilde g_{ba} &=   \frac{
    {_F\langle}\Psi| c_b ^\dagger c_a|\Phi{ \rangle _F}
    }
       {
         {_F\langle}\Psi|\Phi{ \rangle _F}
       }
       =
\big[\Phi (\Psi^\dagger \Phi) ^{-1} \Psi ^\dagger \big]_{ab}       
\end{align*} 
Similarly the 4-point correlation function is evaluated as
\begin{align*} 
g^U_{ba;cd} &= 
{_F\langle} \Psi | c_b c_a  c_c ^\dagger  c_d ^\dagger | \Phi {\rangle _F}
\\
&= 
 \det\nolimits_{M +2}
\mvec
{e_{ab} ^\dagger }
{\Psi ^\dagger }
(e_{cd},\Phi), \quad
e_{ab}  = (e_a,e_b),\
e_{cd} = (e_c,e_d)
\\
&= 
\det\nolimits_{2}[ e_{ab} ^\dagger e_{cd}
-
e_{ab} ^\dagger \Phi (\Psi^\dagger \Phi) ^{-1} \Psi ^\dagger e_{cd}
]
\det\nolimits_{M} (\Psi ^\dagger \Phi)
\\
&= 
\det\nolimits_{2}[ e_{ab} ^\dagger \big[
I_N
-
 \Phi (\Psi^\dagger \Phi) ^{-1} \Psi ^\dagger \big] e_{cd}
]
\det\nolimits_{M} (\Psi ^\dagger \Phi)
\\
&= 
\det\nolimits_{2}\bigg[
\mvec
{e_a ^\dagger }{e_b ^\dagger }
\big[I_N-
\Phi (\Psi^\dagger \Phi) ^{-1} \Psi ^\dagger 
\big]
({e_c },{e_d  })
\bigg]
\det\nolimits_{M} (\Psi ^\dagger \Phi)
\end{align*} 

It implies the Wick theorem
\begin{eqnarray*} 
  g_{ba;cd} &\equiv&
  \frac{
    {_F\langle}\Psi|c_b c_a c_c ^\dagger c_d ^\dagger |\Phi{ \rangle _F}
    }
       {
         {_F\langle}\Psi|\Phi{ \rangle _F}
       }
=  g^U_{ab;cd} /\det\nolimits_M (\Psi|\Phi)
\\
&=& 
\det\nolimits_{2}
\mvec
{e_a ^\dagger }{e_b ^\dagger }
g
({e_c },{e_d  })
\\
&=& 
g_{ac}g_{bd}
-
g_{ad}g_{bc}.
\end{eqnarray*} 
A generic case for the $2M$-point function follows straightforwardly as
\begin{align*} 
g_{
a_M\cdots a_2 a_1;
b_1b_2\cdots b_M
}
& \equiv 
  \frac{
    {_F\langle}\Psi|
    c_{a_M}\cdots c_{a_1}c ^\dagger _{b_1}\cdots c ^\dagger _{b_M}
    |\Phi{ \rangle _F}
    }
       {
         {_F\langle}\Psi|\Phi{ \rangle _F}
       }
\\
&=
\det\nolimits_{M}
e_{a_1,\cdots,a_M} ^\dagger 
g
e_{b_1,\cdots,b_M}
\\
&= \det\nolimits_{M} g^{R},
\end{align*} 
where 
$
  e_{i_1,\cdots,i_M} = (e_{i_1},\cdots,e_{i_M})\in M(N,M)
$ and
$ \{g^{R}\}_{i,j} =  g_{a_ib_j},\ (i,j=1,\cdots,M)
$.
  }



\end{document}

%% file: mydef.tex
\newcommand{\acmt}[2]{{\{}#1,#2{\}}}

\newcommand{\eas}[0]{\begin{eqnarray*}}
\newcommand{\eae}[0]{\end{eqnarray*}}
\newcommand{\les}[0]{\begin{equation}}
\newcommand{\lee}[0]{\end{equation}}
\newcommand{\leas}[0]{\begin{eqnarray}}
\newcommand{\leae}[0]{\end{eqnarray}}
\newcommand{\mchss}[4]
{
\left\{
\begin{array}{cc}
#1 & #2   \\
#3 & #4
\end{array}
\right.
}

\newcommand{\mmat}[4]
{
\left(
\begin{array}{cc}
#1 & #2 \\
#3 & #4 
\end{array}
\right)
}

\newcommand{\mvec}[2]
{
\left(
\begin{array}{c}
#1  \\
#2  
\end{array}
\right)
}

\newcommand{\mvecthree}[3]
{
  \left(
    \begin{array}{c}
#1  \\
#2  \\
#3  
    \end{array}
    \right)
}